\documentclass{PoS}
\usepackage{textcomp}
\usepackage{url}

\title{A pointing solution for the medium size telescopes for the Cherenkov Telescope Array}

\ShortTitle{A pointing solution for the MST for CTA}

\author{\speaker{D. Tiziani}\\
        Erlangen Centre for Astroparticle Physics\\
        Erwin-Rommel-Str. 1, 91058 Erlangen, Germany\\
        E-mail: \email{domenico.tiziani@fau.de}}

\author{M. Garczarczyk\\
        Deutsches Elektronen Synchrotron\\
        Platanenallee 6, 15738 Zeuthen, Germany\\
       E-mail: \email{markus.garczarczyk@desy.de}}
       
\author{L. Oakes\\
	Institut f\"ur Physik, Humboldt-Universit\"at zu Berlin\\
	Newtonstr. 15, 12489 Berlin, Germany\\
	E-mail: \email{loakes@physik.hu-berlin.de}}
	
\author{U. Schwanke\\
	Institut f\"ur Physik, Humboldt-Universit\"at zu Berlin\\
	Newtonstr. 15, 12489 Berlin, Germany\\
	E-mail: \email{schwanke@physik.hu-berlin.de}}
	
\author{C. van Eldik\\
        Erlangen Centre for Astroparticle Physics\\
        Erwin-Rommel-Str. 1, 91058 Erlangen, Germany\\
        E-mail: \email{Christopher.van.Eldik@physik.uni-erlangen.de}}

\author{for the CTA Consortium}

\abstract{The pointing capability of a telescope in the Cherenkov Telescope Array (CTA) is a crucial aspect in the calibration of the instrument. It describes how a position in the sky is transformed to the focal plane of the telescope and allows precise directional reconstructions of atmospheric particle showers.
The favoured approach for pointing calibrations of the Medium Size Telescopes (MST) is the utilisation of an CCD-camera installed in the centre of the dish, which images the night sky and the focal plane simultaneously. 
The technical implementation of this solution and test results taken over a period of one year at the MST prototype in Berlin/Adlershof are presented. Investigations of pointing calibration precision with simulated data and real data taken during test runs of the prototype telescope will also be shown.
}

\FullConference{35th International Cosmic Ray Conference - ICRC2017-\\
		10-20 July, 2017\\
		Bexco, Busan, Korea}

\begin{document}

\section{Introduction}

CTA, the next generation ground based gamma-ray telescope, is currently in a pre-construction phase. It will consist of three different sizes of Imaging Atmospheric Cherenkov Telescopes (IACT) to detect Cherenkov light from gamma-ray induced air-showers in a wide energy range. This light is collected by mirror dishes and projected onto specialized Cherenkov cameras. \\
For the reconstruction of the arrival direction of the original gamma-ray photon it is important to know the exact orientation of the telescope with respect to the sky and the alignment of the Cherenkov camera with respect to the telescope's optical axis at the time of its detection. This geometric configuration is also called the pointing of a telescope. The favoured approach for the calibration of the pointing of the MST uses an optical CCD-camera (the SingleCCD) that is mounted in the centre of the mirror dish of the telescope and images the Cherenkov camera and the night sky around it simultaneously. The Single-CCD concept has also been successfully tested by the H.E.S.S. collaboration \cite{lennarz}.

\section{Single-CCD Concept}

The pointing calibration is performed in two steps. In the first step, the telescope is pointed at different positions in the night sky during special pointing runs. In these runs, a screen in front of the Cherenkov camera is used to see light spots from bright stars. These spots are captured by the SingleCCD, and their positions on the screen for a given telescope alignment are used for the derivation of a pointing model for the telescope. This technique has been used successfully e.g. by the H.E.S.S. collaboration \cite{gillessen}. The second step is performed during data taking, while the Cherenkov camera is operative. The SingleCCD now images the spots of LEDs installed on the Cherenkov camera housing surrounding the centre of the focal plane and the stars behind the Cherenkov camera. The positions of the LEDs can be used to determine the position of the Cherenkov camera with respect to the optical axis while the positions of the stars can be used to deduce the orientation of the telescope on the sky. With this information the previously derived pointing model can be refined and short term deviations can be corrected.\\
An alternative approach utilizes two separate CCD-cameras: A LidCCD, which images reflected star spots and positioning LEDs on the Cherenkov camera, and a SkyCCD which tracks star positions for pointing corrections during data taking. The downsides of this method are an increased complexity and uncertainties in the stability of the relative orientation between the two cameras. The advantage is a possible increase of precision in the determination of the pointing due to a smaller field of view of the SkyCCD compared to the SingleCCD.  

\section{Camera Hardware}
For the SingleCCD camera an Apogee ASPEN CG8050 with electronic shutter, chip dimensions 3296 $\times$ 2472 pixels, and pixel size 5.5\,\textmu m is used. With a 50\,mm f/1.8 Nikkor lens, images with a field-of-view of 20.5 $\times$ 15.5\,deg$^2$ can be taken. In these images, a pixel diameter corresponds to about 22'' in the sky. The CCD chip of the camera can be cooled to a constant temperature via an integrated Peltier element to avoid thermal expansion and to preserve a constant imaging geometry. The camera data is read out via a standard Ethernet interface.\\
\begin{figure}
  \centerline{{\includegraphics[width=0.8\textwidth]{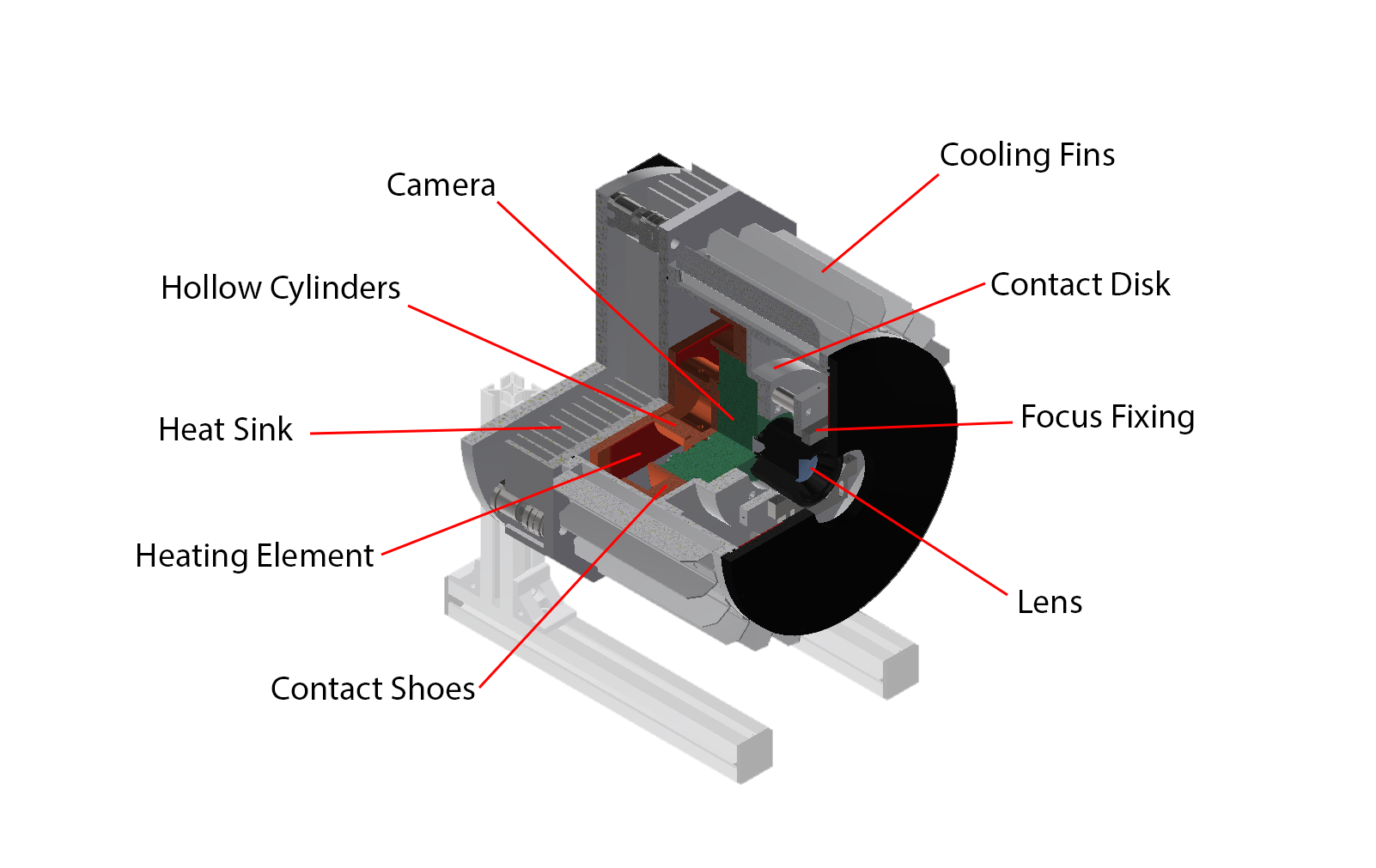}}}
  \caption{Schematic view of the SingleCCD camera in its casing. The rigid support structure (focus fixing) and heat conducting elements (contact disk, contact shoes, cooling fins, heat sink, and hollow cylinder) are highlighted. The heating element for the heat sink is also shown.}
  \label{fig:housing}
\end{figure}

The camera is mounted inside a custom-made casing that shields the electronics and optics from environmental impacts (see Figure \ref{fig:housing}). The housing is fully IP67 compliant and connects the SingleCCD rigidly with the telescope dish. It comprises several heat conducting structures that grant a good heat dissipation from the camera to the environment. A heating element is installed to the front of the casing to keep the front window free of ice at low temperatures. Another heating element at the backside of the housing assures that the heat sink temperature stays above the desired CCD temperature as the camera internal electronics do not support chip heating.\\ This is the second version of the casing prototype. It is currently in construction and will replace the currently used version, which has been introduced in \cite{gamma16}, at the prototype telescope.

\section{Precision Studies}
\label{sec:precision}
\begin{figure*}[t!]
\centering
\includegraphics[width=0.4\textwidth]{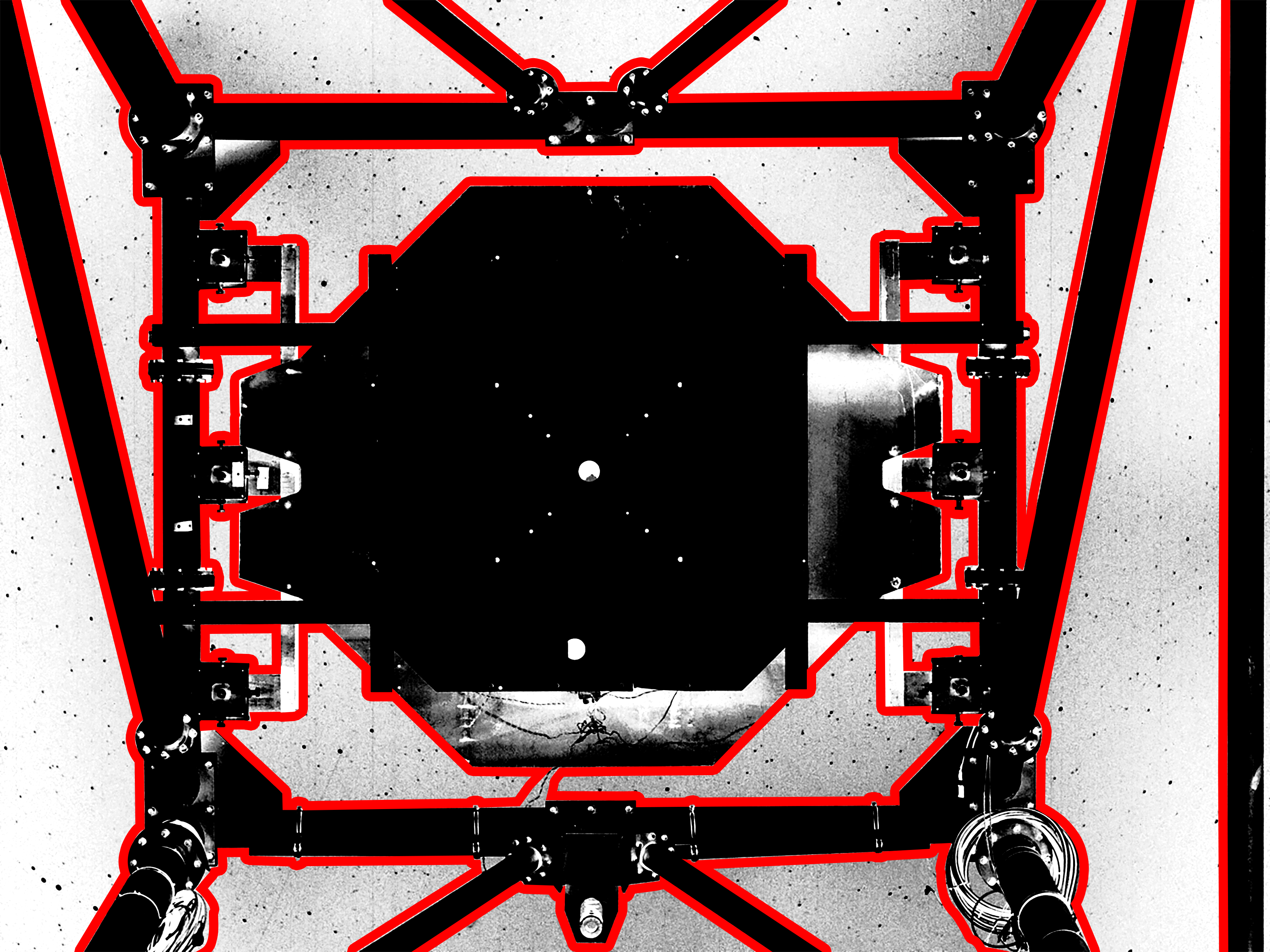}
\includegraphics[width=0.5\textwidth]{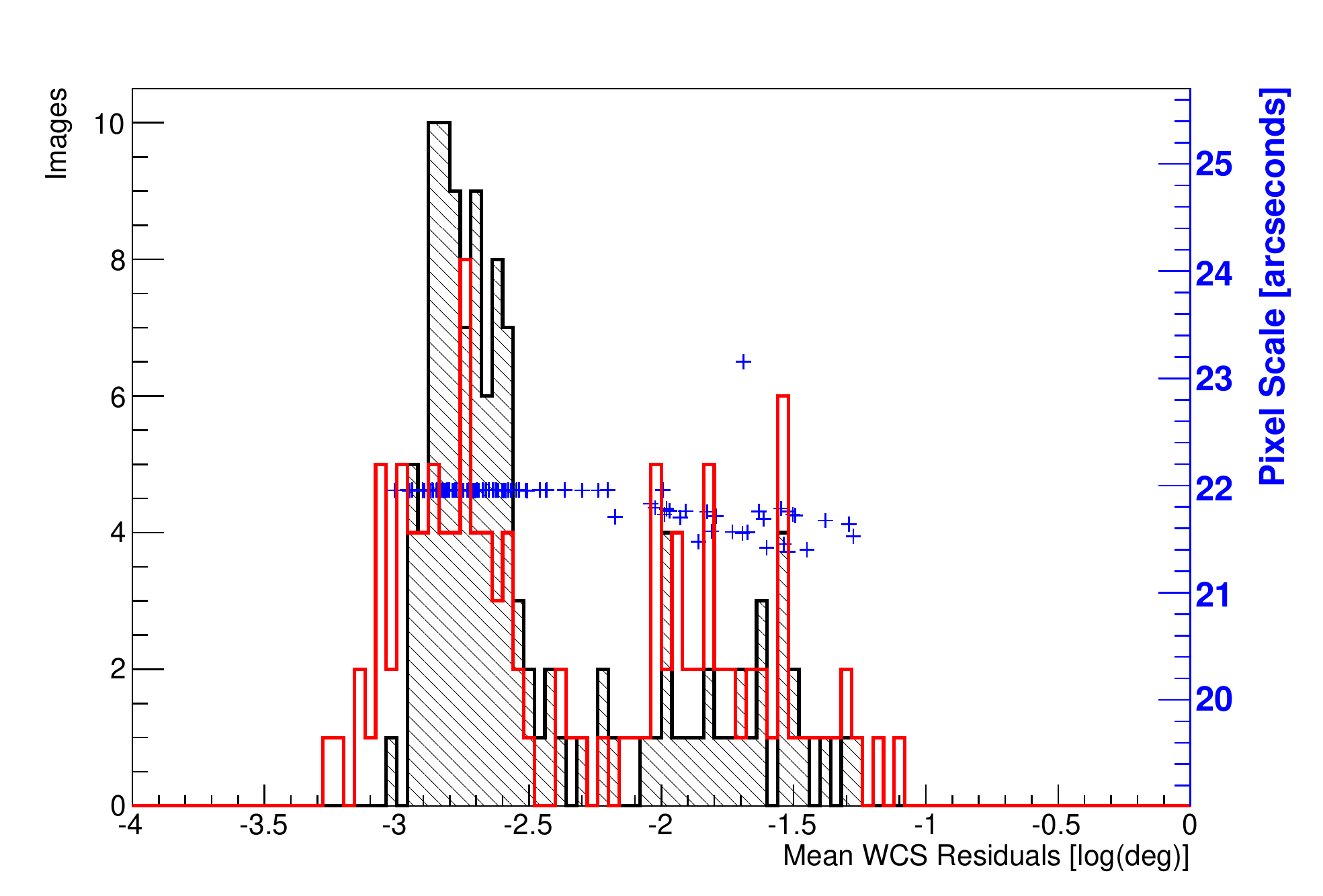}
\caption{({\bf Left:}) SingleCCD image taken at the MST prototype in Berlin Adlershof. Colors are inverted. Red contour marks the area which is obstructed by the telescope structure. ({\bf Right:}) Distributions of mean WCS residuals for images taken at the MST prototype (black histogram) and simulated images (red histogram). Blue plus signs indicate the corresponding pixel scales. See text for explanations.\label{fig:mask}
}
\end{figure*}

The images taken by the SingleCCD are analysed with the software {\tt astrometry.net} \cite{lang}. This calibration software calculates a world coordinate system (WCS) for a given image of the night sky. This information is then used to derive the absolute pointing position of the telescope. One challenge of the SingleCCD approach is that a large area in the pointing images is occupied by the Cherenkov telescope camera and its support structure, in particular the central part of the images, where the pointing direction of the telescope is positioned (see Figure \ref{fig:mask} left). The fact that no stars can be detected in large areas of the images separates this application of {\tt astrometry.net} from its standard use case.  In order to determine the impact of this shadowing effect on the precision of the pointing measurement, a simulation software for SingleCCD images has been developed. 
\\
In \cite{gamma16} it has been shown that the pointing of simulated images shadowed by a square-shaped Cherenkov camera can be reconstructed to a precision of $\sim 1$ arcsecond. The simulation is now used to reproduce images that have been taken in a test run during one night at the prototype telescope with the full camera support structure. Both real images and simulated images are analysed with {\tt astrometry.net}. For each identified star in an image, the residual angular difference between expected position and calculated WCS position is determined. In Figure \ref{fig:mask} right, the distribution of the mean residual is shown both for simulations and for real images. The very good correspondence between simulation and real data indicates that simulation and imaging hardware work as expected and that effects like distortions which are not implemented in the simulation do not affect the calibration of real images. It can be seen that there are two separate classes of solutions. For about 25\% of the image calibrations, the residuals become quite large (> 0.5'). These are suspected to originate from local minima that are reached by the WCS fit routine. The larger class of calibrations show mean residuals in the range of a few arcseconds (< 20''). The two classes can also be discriminated via the derived pixel scale of the WCS. This scale specifies the angular correspondence of a pixel diameter in an image. While for the higher precision calibrations, this quantity is close to the correct value, it fluctuates strongly for the lower precision fits.\\
These two classes, which also show up for simulated images, indicate that there are still some possibilities to optimize the image calibration procedure. It is currently under investigation how the WCS fit can be improved further. Moreover, the order of magnitude of the observed mean WCS residuals is equal to that of the residuals of first derived pointing models  (cf.~Sect.~\ref{sec:pointing}) which can be taken as a hint that both originate from the same imprecision in the fit.

\section{Pointing Models}
\label{sec:pointing}
For MSTs the pointing precision needs to be such that any position in the focal
plane can be mapped to a sky position with an accuracy of better than $7^{''}$ (space angle). To stay
within this error budget the pointing accuracy of the SingleCCD alone must be better than 
$7^{''}$. Using data recorded at the MST prototype in Berlin, pointing models for the SingleCCD have 
been derived in order to study the resulting precision and the time evolution of the model
parameters.

The input data for the calculation of pointing models were obtained in clear
nights where the MST tracked about 100 sky positions at elevations greater than $45^{\circ}$.
The observations were conducted in a robotic fashion and took several hours since
2.5\,min were needed to slew to a new position and to track it while the SingleCCD
was read out. The exposure of the SingleCCD was set to 20\,s; the resulting images
were stored in FITS format along with the time evolution of the azimuth and elevation
targeted by the drive system (az$_D$, el$_D$). All measurements were conducted
in direction of either continuously increasing or decreasing elevation to test for
hysteresis effects.

The images were solved using the {\tt astrometry.net} software to find the 
equatorial coordinate of the center of the SingleCCD FoV; the 90\% of the images
for which a good-quality solution (cf.~Sect.~\ref{sec:precision}) could be found 
were then used to derive a pointing model. To this end, the time $t_0$ corresponding
to the centre of the exposure interval was estimated and used to derive 
the coordinate targeted by the drive system (az$_D(t_0)$,el$_D(t_0)$). In a similar
way, the found equatorial coordinate was converted to obtain the 
azimuth and elevation (az$_C(t_0)$, el$_C(t_0)$) that corresponded to 
the center of the FoV of the SingleCCD at time $t_0$. A pointing model is defined
by describing the difference in elevation and azimuth, 
$$
\mbox{el}_D - \mbox{el}_C = f_1(\mbox{el}_C,\mbox{az}_C,\vec{q}) \mbox{\ \ and\ \ } 
\mbox{az}_D - \mbox{az}_C = f_2(\mbox{el}_C,\mbox{az}_C,\vec{q}) \mbox{,}
$$
by suitable functions $f_1$ and $f_2$ that depend on a vector of adjustable
parameters ($\vec{q}$). The best parameter values were found by minimizing 
the angular distance between the predicted and the measured 
az$_D$ and el$_D$ positions over all images recorded in one night.

\begin{figure*}[t!]
\centering
\includegraphics[width=0.33\textwidth]{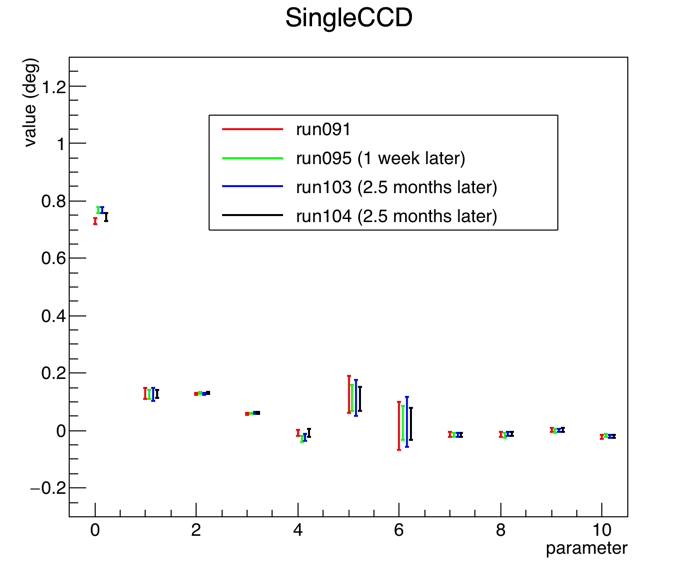}
\includegraphics[width=0.66\textwidth]{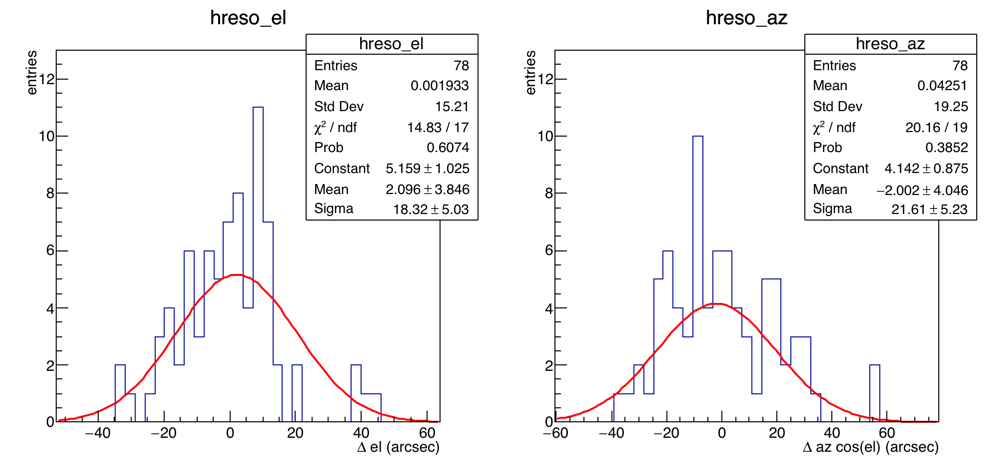}
\caption{Results of pointing studies for a SingleCCD mounted on the MST prototype 
in Berlin Adlershof. ({\bf Left:}) Time evolution of the 11 parameters of the 
fitted pointing models. The parameters values (in degrees) are shown vs
the parameter number ($0,\ldots,10$). The colors denote four different parameter
sets derived for nights in a period of 2.5 months. Note that the 
values have been slightly shifted horizontally for better visibility.
({\bf Middle, Right:}) Distributions of the residuals between measured and predicted elevation and
azimuth positions. See text for explanations.\label{fig:pointing}
}
\end{figure*}

The employed pointing model had 11 parameters, the most basic ones being 
constant offsets in azimuth and elevation along with tilts of the azimuth
axis in the East-West ($\Theta_{EW}$) and the North-South ($\Theta_{NS}$) direction, respectively. Figure
\ref{fig:pointing} (left) shows the time evolution of the 11 parameters 
over a period of 2.5 months. It is evident that the parameter values are fairly
stable; it was also found that the SingleCCD shows no sagging with elevation 
and that the values of the tilt parameters ($\Theta_{EW}$ and $\Theta_{NS}$) agree with 
values derived for a SkyCCD that exclusively observes stars in the sky.
Figure \ref{fig:pointing} also presents the residuals between model and prediction
in elevation (middle) and azimuth (right). Both distribution are centred at
zero, demonstrating that model describes the data on average. The width of the two 
distributions is, however, at the level of $20^{''}$. This implies that the 
current per-axis error exceeds the MST pointing precision that is aimed for.
It is currently under study in how far the broadening of the residuals is 
due to environmental effects (e.g.~the enhanced night-sky background levels
in Berlin Adlershof), the telescope structure and drives (e.g.~vibrations
and tracking deviations), the SingleCCD hardware, or the analysis procedure
(e.g.~the determination of equatorial coordinates by the {\tt astrometry.net} 
software).

\section{Conclusion}
Results from one year of testing at the MST prototype show that the SingleCCD pointing solution can be used for calibrations and to derive pointing models. The residuals of the calculated pointing models imply that the desired precision is not yet fully reached. Detailed studies give a hint that the algorithm of the used {\tt astrometry.net} WCS fit routine might be one reason for this imprecision but environmental influences e.g. from the high night sky background in Berlin can not be ruled out yet. Further tests are still needed to improve the method and to show that it can be applied in the final state for the MSTs in CTA.

\section*{Acknowledgments}
This work was conducted in the context of the CTA MST Structure Working Group. We gratefully acknowledge financial support from the agencies and organizations listed here:\\ \url{http://www.cta-observatory.org/consortium_acknowledgments}.\\This work was supported by the German Ministry of Education and Research under grant identifier 05A14WE2.

\end{document}